\documentclass[prl,a4paper,twocolumn,superscriptaddress,longbibliography]{revtex4-2}

\usepackage{amssymb}
\usepackage{amsmath}
\usepackage{amsfonts}
\usepackage{graphicx}
\usepackage{bm}
\usepackage{xcolor}
\usepackage{multirow}
\usepackage{comment}
\usepackage{pgfplots}
\usepackage{tikz}

\usepackage[colorlinks,citecolor=blue,linkcolor=red,urlcolor=blue]
{hyperref}

\DeclareMathOperator{\sgn}{sgn}

\makeatletter
\newcommand\footnoteref[1]{\protected@xdef\@thefnmark{\ref{#1}}\@footnotemark}
\makeatother

\begin{document}

\title{Chirality inversion and radius blow-up of a N\'eel-type skyrmion by a Pearl vortex}

\author{S. S. Apostoloff}

\affiliation{\mbox{L. D. Landau Institute for Theoretical Physics, Semenova 1-a, 142432, Chernogolovka, Russia}}

\author{E. S. Andriyakhina}

\affiliation{Moscow Institute for Physics and Technology, 141700 Moscow, Russia}

\affiliation{\mbox{L. D. Landau Institute for Theoretical Physics, Semenova 1-a, 142432, Chernogolovka, Russia}}

\author{P. A. Vorobyev} 

\affiliation{School of Physics, The University of New South Wales, Sydney 2052, Australia}

\author{Oleg A. Tretiakov}

\affiliation{School of Physics, The University of New South Wales, Sydney 2052, Australia}

\author{I. S. Burmistrov}

\affiliation{\mbox{L. D. Landau Institute for Theoretical Physics, Semenova 1-a, 142432, Chernogolovka, Russia}}

\affiliation{Laboratory for Condensed Matter Physics, HSE University, 101000 Moscow, Russia}

\begin{abstract}
We develop a theory for the coaxial configuration of a N\'eel-type skyrmion and a Pearl vortex in thin superconductor-chiral ferromagnetic heterostructures. 
Using direct numerical solution of the Euler-Lagrange equation and micromagnetic simulations we demonstrate that the inhomogeneous magnetic field of the Pearl vortex significantly modifies the skyrmion profile with respect to the one in the absence of the vortex. 
We discover drastic enlargement of the skyrmion's radius and inversion of the skyrmion's chirality. To unravel physics behind these effects we invent 
novel two-parameter ansatz for the magnetization profile of the skyrmion in the presence of the vortex. 
Chirality inversion and radius blow-up are controlled not only by the material parameters of the heterostructure but also by the thickness of the superconductor. Our findings can have implications for Majorana modes localized at skyrmion-vortex pairs. 
\end{abstract}

\date{\today}

\maketitle


An interest to coexistence of magnetism and superconductivity in heterostructures has resurgence last two decades~\cite{Ryazanov2004,Lyuksyutov2005,Buzdin2005,Bergeret2005,Eschrig2015}.
In particular, superconductor--chiral ferromagnet (SF) bilayers have recently attracted much attention \cite{Back2020,Gobel2021,Zlotnikov} as systems 
hosting two topologically nontrivial configurations: skyrmions stabilized by Dzyaloshinskii--Moriya interaction (DMI) in a ferromagnetic film  \cite{Bogdanov1989} and vortices in a superconductor. 
Skyrmions in SF bilayers demonstrate rich physics 
 inducing Yu-Shiba-Rusinov-type bound states \cite{Pershoguba2016,Poyhonen2016}, modifying the Josephson effect \cite{Yokoyama2015}, and changing the superconducting critical temperature \cite{Proshin2022}. Skyrmion-vortex pairs host Majorana modes \cite{Chen2015,Yang2016,Gungordu2018,Mascot2019,Rex2019,Garnier2019,Rex2020,Gungordu2022} and can serve as a scalable topological quantum computing platform \cite{Nothhelfer2022}. Experimental demonstration of stable skyrmion-vortex coexistence has been recently reported in [Ir$_1$Fe$_{0.5}$Co$_{0.5}$Pt$_1$]$^{10}$/MgO/Nb sandwich \cite{Petrovic2021}.

Skyrmions and vortices in SF bilayers can form bound pairs due to interplay of spin-orbit coupling and proximity effect \cite{Hals2016,Baumard2019} as well as due to interaction via stray fields \cite{Dahir2019,Menezes2019,Dahir2020,Andriyakhina2021}. Traditionally, analysis of Majorana modes in skyrmion-vortex pairs ignores the effect of stray fields. However, for a thin SF bilayer the interaction due to stray fields result in dramatic effect: repulsion of a N\'eel--type skyrmion from a Pearl vortex to a finite distance stable position \cite{Andriyakhina2021}.

In this Letter we develop a theory for coaxial configuration of a N\'eel-type skyrmion and a Pearl vortex in a thin SF heterostructure. We use two complementary approaches: direct numerical solution of the Euler-Lagrange equation, cf. Eq.~\eqref{eq:ELE_theta_coax} and micromagnetic simulations based on Landau-Lifshitz-Gilbert equation. To perform the free energy minimization we invent a two-parameter ansatz, cf. Eq.~\eqref{eq:ansatz}, inspired by 
numerical solutions of the Euler-Lagrange equation.
We create this anzatz based on naturally arising synergy of a sole skyrmion's profile and the magnetization induced by the vortex.
We 
find that the inhomogeneous magnetic field of the Pearl vortex significantly modifies the skyrmion's profile  and 
results in two effects: (i) drastic enlargement of skyrmion's radius with respect to its radius in the absence of the vortex and (ii) inversion of the chirality with respect to the `natural' chirality fixed by the sign of DMI. Both effects can have implications for existence of Majorana modes localized at skyrmion-vortex pairs. 
 
\noindent\textsf{\color{blue} Model.} 
We consider a heterostructure consisting of two films, superconducting and ferromagnetic, separated by the thin insulating layer that suppresses the proximity effect. The superconducting film is supposed to be much thinner than the London penetration depth, $d_S{\ll}\lambda_L$, and contain a Pearl vortex. The main goal of our paper is to study a N\'{e}el-type skyrmion in the ferromagnetic film located coaxially on the top of a Pearl vortex situated in the superconducting film. 

The free energy of a thin chiral ferromagnetic film interacting with a Pearl vortex is given by
\begin{align}
	\mathcal{F}[\bm{m}]  = & d_F \!\int d^2\! \bm{r} \bigl \{ A (\nabla \bm{m})^2 + K(1- m_z^2) + D [m_z \nabla \cdot \bm{m}
	\notag \\
	& - (\bm{m}\cdot \nabla) m_z ] - M_s \bm{m}\cdot \bm{B}_{\rm V}|_{z=+0} \bigr \} .
	\label{eq:MagFe}
\end{align}
Here $\bm{m}(\bm{r})$ is the unit magnetization vector, $M_s$ is the saturation magnetization, and $d_F$ is the thickness of the ferromagnetic film. Parameters $A{>}0$, $K{>}0$, and $D$ stand for the exchange, perpendicular anisotropy, and DMI constants, respectively. The $z$ axis is directed perpendicular to the film.  The magnetic field due to the Pearl vortex, ${\bm B}_{\rm V}$, is centered at the origin,
\begin{eqnarray}
	{\bm B}_{\rm V}  = \phi_0 \sgn(z) \nabla 
	\int \frac{d^2\bm{q}}{(2\pi)^2} \frac{e^{-q |z| +i \bm{q}\bm{r}}}{q(1+2q\lambda)},
	\label{eq:vortex:B}
\end{eqnarray}
where $\phi_0=h c/2e$ is the flux quantum and $\lambda=\lambda_L^2/d_S$ is the Pearl length \cite{Pearl1964}.
The free energy~$\mathcal{F}[\bm{m}]$ is normalized in such a way that $\mathcal{F}=0$ for the ferromagnetic state, $m_z{=}1$, in the absence of the Pearl vortex, ${\bm B}_{\rm V}{=}0$.

The magnetization of the N\'{e}el-type skyrmion coaxial with the Pearl vortex, due to the radial symmetry of the problem, can be sought as
$\bm{m} {=} \bm{e}_r \sin \theta (r)  {+} \bm{e}_z \cos \theta (r).$ 
Minimizing the free energy~$\mathcal{F}[\bm{m}]$ with respect to the skyrmion angle~$\theta(r)$, one can derive the Euler-Lagrange equation, 
\begin{gather}
	\frac{\ell_{w}^{2}}{r} \partial_r \big[r \partial_r \theta(r)\big]
	-\frac{(\ell_{w}^{2}+r^2)}{2 r^2}\sin2 \theta(r)    
	+ 2\epsilon \frac{ \sin^2 \theta(r)}{r/\ell_{w}} 
	\notag\\
	+\gamma  [b_z(r)\sin\theta(r)-b_{r}(r)\cos\theta(r)] = 0,
	\label{eq:ELE_theta_coax}
\end{gather}
where ${\ell_{w} {=} \sqrt{A/K}}$ is the domain wall width. 
We introduced two dimensionless parameters: the DMI strength ${\epsilon{=}D/2\sqrt{AK}}$ 
and the effective strength of the Pearl vortex ${\gamma{=}(\ell_{w}/\lambda)(M_s\phi_0/8\pi A)}$.
The functions $b_r(r)$ and $b_z(r)$ are the rescaled projections of the  magnetic field of the Pearl vortex in the ferromagnetic film, ${{\bm B}_V|_{z=+0}{=}  
{-}(\phi_0/4\pi \ell_{w}\lambda) [b_r(r)\bm{e}_r{+}b_z(r){\bm e}_z]}$.

Equation~\eqref{eq:ELE_theta_coax} should be supplemented by appropriate boundary conditions. One of them, $\theta(r{\to} \infty){=}0$, indicates that the magnetization far from the origin, where the vortex and the skyrmion are situated, is uniform, i.e. $m_z{=}1$. The appropriate choice of the second boundary condition at $r{=}0$ determines the specific system configuration that we describe in detail below. 

\noindent\textsf{\color{blue}Vortex without skyrmion.} 
If condition $\theta(r{=}0){=}0$ is 
assumed, the solution of Eq.~\eqref{eq:ELE_theta_coax}, ${\theta(r){=}\theta_{\gamma}(r)}$ describes the magnetization of initially homogeneous ferromagnetic film without a skyrmion in the magnetic field of the Pearl vortex. In this paper we focus on the most realistic case 
in which the Pearl length $\lambda$ is much larger than the skyrmion radius. Then it is 
enough to 
consider the rescaled magnetic field of the vortex  
as well as the magnetization angle ${\theta(r)}$ at distances $r{\ll}\lambda$ only. In that approximation $b_r(r){\approx} b_z(r){\approx}\ell_{w}/r$, and the solution of Eq.~\eqref{eq:ELE_theta_coax} with the boundary condition $\theta(r{=}0){=}0$ can be found in the analytical form for not too large vortex strength~\footnote{To derive the analytical result of Eq.~\eqref{eq:thetaV}, we assume $\theta_{\gamma}(r)$ to be small and linearize Eq.~\eqref{eq:ELE_theta_coax}. The maximal absolute value of $\theta_{\gamma}(r)$ equals approximately $0.4\gamma$. Therefore, Eq.~\eqref{eq:thetaV} is correct provided $\gamma{\lesssim1}$.},
\begin{equation}
	\theta_{\gamma}(r)\approx \gamma \left[K_1(r/\ell_{w})-\ell_{w}/r\right],
	\quad \gamma\lesssim1,
	\label{eq:thetaV}
\end{equation}
where $K_1(x)$ is the modified Bessel function of the second kind.

\noindent\textsf{\color{blue}Skyrmion with vortex.} 
If we impose condition $\theta(r{=}0){=}\chi\pi$ with $\chi{=}{\pm}1$,
the solution of Eq.~\eqref{eq:ELE_theta_coax} corresponds to a skyrmion with chirality $\chi$. The chirality $\chi{=}{+}1$ (${-}1$) means that the in-plane projection of magnetization is directed from (to) the center of the skyrmion. 
As well-known~\cite{Bogdanov1994}, without the Pearl vortex, $\gamma{=}0$, Eq.~\eqref{eq:ELE_theta_coax} has solution for a single chirality, ${\chi{=}\sgn(\epsilon)}$, and only for $|\epsilon|{<}2/\pi$. The ferromagnet with $|\epsilon|{>}2/\pi$ is not in homogeneous, but in spiral, state so we do not study such strong DMI. It should be emphasized that in the presence of the vortex, $\gamma{>}0$, solutions of Eq.~\eqref{eq:ELE_theta_coax} with 
single or both chiralities $\chi{=}{\pm}1$ can be found depending on the magnitudes of $\gamma$ and $\epsilon$.

The solution of Eq.~\eqref{eq:ELE_theta_coax} with the boundary condition $\theta(r{=}0){=}\chi\pi$ can be found numerically, e.g. by the ``shooting'' method. However, this procedure is computationally expensive, because the ``shooting'' parameter should be sought with exponentially high accuracy with respect to the accuracy of the solution itself. Nevertheless, there is an alternative way to find an approximate solution, which is convenient both for numerical computations and analytical study. 

\noindent\textsf{\color{blue}Skyrmion-vortex ansatz.} For the description of the skyrmion-vortex coaxial pair we propose to use the following ansatz, 
\begin{equation}
	\label{eq:ansatz}
	\theta_{R,\delta,\gamma}(r) =\theta_{R,\delta}(r)+ \theta_{ \gamma}(r)\cos\theta_{R,\delta}(r), 
\end{equation} 
which is a modified version of the well-known {360\textdegree} domain-wall ansatz,
\begin{equation}
	\theta_{R,\delta}(r) = 2\arctan\dfrac{\sinh(R/\delta)}{\sinh(r/\delta)}.
\end{equation}
Then one should insert $\theta_{R,\delta,\gamma}(r)$ instead of $\theta(r)$ into the free energy $\mathcal{F}[\bm{m}]$ and minimize it with respect to two parameters only:
$R$ and $\delta$. While the latter plays a role of the skyrmion wall width, the former encodes both the skyrmion chirality, $\chi{=}\sgn R$, and its radius $|R|$.

The qualitative idea of the construction of ansatz~\eqref{eq:ansatz} is as follows. One may expect, even in the presence of the magnetic field of the Pearl vortex, that the skyrmion shape is approximately described by the {360\textdegree} domain wall ansatz, i.e. ${\theta(r){\approx}\theta_{R,\delta}(r){+}\delta\theta(r)}$. To determine $\delta\theta(r)$, we consider different areas of the ferromagnet, near the center of the skyrmion, 
at $r{\sim} |R|$, and far from the origin. Far from or near the center of the skyrmion the magnetization is nearly homogeneous, ${m_z{\approx}\pm1}$, and its variation is determined mostly by the magnetic field of the vortex, hence ${\delta\theta(r){\approx} {\pm}\theta_{\gamma}(r)}$. In the intermediate region,  
${m_z{\approx}\cos\theta_{R,\delta}(r)}$, so it is natural to smooth out $\delta\theta(r)$ as ${\delta\theta(r){\approx} \theta_{\gamma}(r)}\cos\theta_{R,\delta}(r)$, and we arrive at ansatz~\eqref{eq:ansatz}.

\noindent\textsf{\color{blue}Micromagnetic modeling.} 
In addition to the direct numerical solution of Eq.~\eqref{eq:ELE_theta_coax} and the minimization of $\mathcal{F}[\bm{m}]$ with the ansatz~\eqref{eq:ansatz}, we have run a series of micromagnetic simulations using OOMMF software \cite{OOMMF}. We simulated the system as a set of classical magnetic vectors placed at the centers of the grid cells in the $xy$-plane. We imposed periodic boundary conditions and used the total energy on such a lattice that transforms into Eq.~\eqref{eq:MagFe} in the continuous limit. 
The Pearl vortex is located at the origin, $x{=}y{=}0$. We initiate the system with the magnetization determined by ansatz~\eqref{eq:ansatz}.

\begin{figure}[!h]
	\includegraphics[width=3.5in]{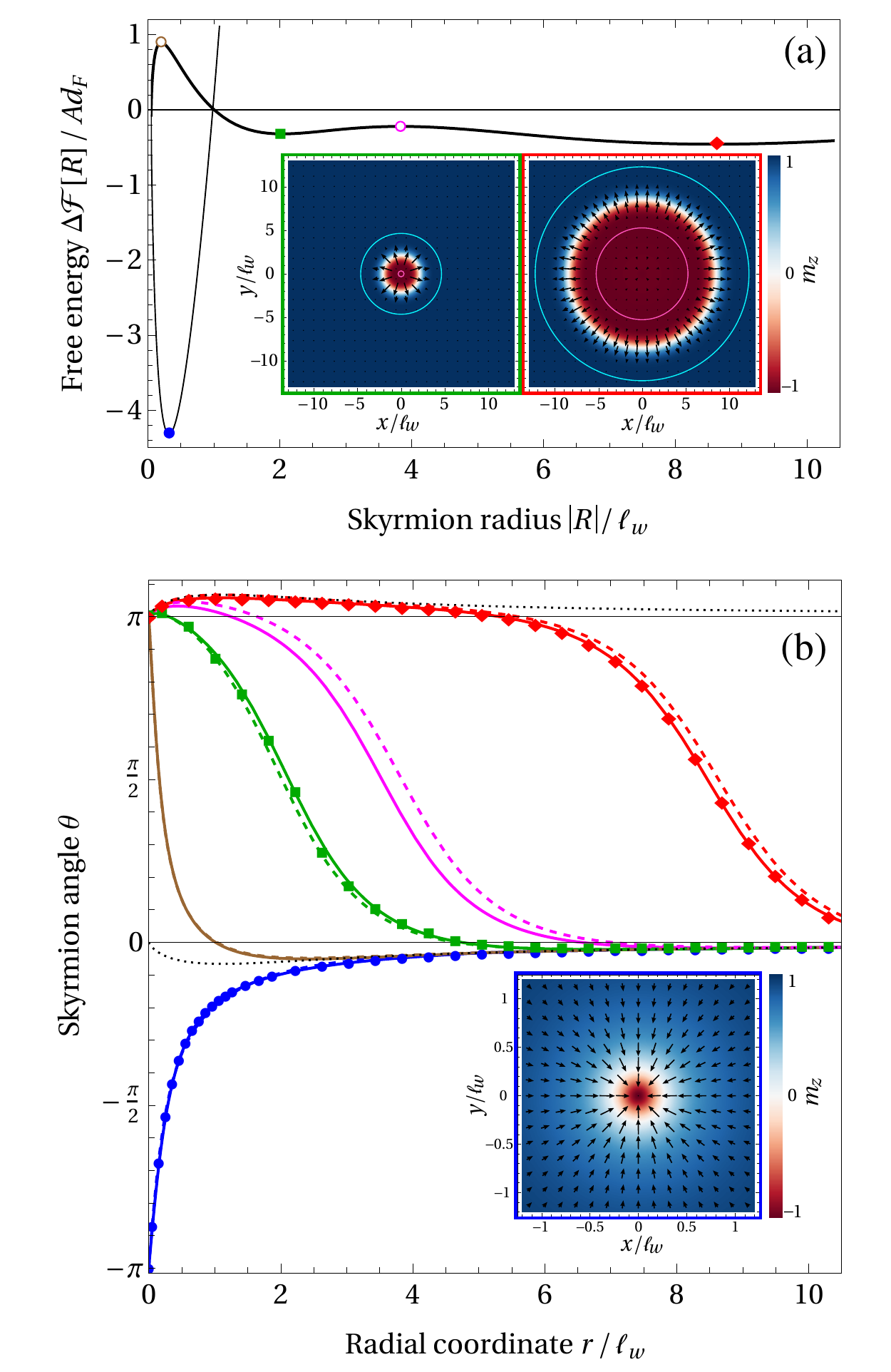}	
	\caption{\textsf{(a)}: Dependence of shifted dimensionless free energy~$ \Delta\mathcal{F}[R]/(A d_F)$ on skyrmion radius~$|R|$ for
	$\epsilon{=}0.3$ and $\gamma{=}0.522$. The thicker (thinner) curve describes it for the positive (negative) chirality, $\chi{=}{+}1({-}1)$. The blue disk, green square, and red diamond indicate the minima of $\Delta\mathcal{F}[R]$, while the brown and magenta circles indicate the maxima. \textsf{(b)}: The skyrmion angles~$\theta(r)$ corresponding to the minima and maxima of $\Delta\mathcal{F}[R]$ from (a). The solid and dashed curves show the exact solution of Eq.~\eqref{eq:ELE_theta_coax} and the approximations given by ansatz~\eqref{eq:ansatz} 
	with $\{R,\delta\}{\approx}\{{-}0.28,0.78\}$, $\{0.16,0.62\}$, $\{2.0,0.91\}$, $\{3.8,0.96\}$, $\{8.6,0.99\}$,
	respectively. The functions $\theta_\gamma(r)$ and $\pi{-}\theta_\gamma(r)$ are plotted as the black dotted curves. The points marked by  blue disks, green squares, and red diamonds are extracted from the micromagnetic modeling. \textsf{The insets}: Spatial distribution of magnetization for stable skyrmion states is
	obtained by micromagnetic simulations. The colors of the inset frames correspond to the color of the curves~$\theta(r)$ of (b).
	The color gradient indicates the magnitude of the $z$-component of magnetization (see color bar). The black arrows guide the magnitude and direction of the in-plane magnetization. The pink and cyan curves shown in the insets in (a) (green and red frames) correspond to the distances from the center at which $m_r{=}0$. 
	}
	\label{fig:FviaR}
\end{figure}

\noindent\textsf{\color{blue}Results.} The magnetic field of the vortex makes the free energy landscape to be more complicated than at $\gamma{=}0$. In particular, there are regions of $\gamma$ and $\epsilon$ in which several minima of $\mathcal{F}[\bm{m}]$ exist. To illustrate such behavior, in Fig.~\ref{fig:FviaR}(a) we plot the shifted free energy~${ \Delta\mathcal{F}[R]{=}\mathcal{F}[R]{-}\mathcal{F}[R{\to}0]}$ normalized by the energy scale~$A d_F$ as a function of the skyrmion radius~$|R|$ for $\epsilon{=}0.3$ and $\gamma{=}0.522$
\footnote{Value $\gamma{=}0.522$ is taken with such a high precision because the skyrmion radius is sensitive to the vortex strength in this range of parameters. Indeed, changing $\gamma$ by 1\% may vary radius $R$ by up to 10\%.}.
Here $\mathcal{F}[R]$ is the free energy~$\mathcal{F}[\bm{m}]$ computed with the help of the ansatz $\theta(r){=}\theta_{R,\delta,\gamma}(r)$ and minimized with respect to the skyrmion wall width~$\delta$ only. As one can see, there are three minima: two minima corresponding to the positive chirality, $\chi{=}{+}1$, (indicated by the green square and red diamond) and one minimum for negative chirality, $\chi{=}{-}1$ (indicated by the blue disk). Potentially, these minima correspond to the skyrmion stable states (see below). We emphasize that the radius of the stable skyrmion configuration indicated by the red diamond in Fig.~\ref{fig:FviaR} is $R{\approx}8.5$ which is approximately 25 times larger than the skyrmion radius $R_0{\approx}0.33$ at $\gamma{=}0$. 
Additionally, $\Delta\mathcal{F}[R]$ has two maxima (indicated by the brown and magenta circles). Physically,
these saddle-like solutions \footnote{We term some solutions of Euler-Lagrange equation~\eqref{eq:ELE_theta_coax} as saddle-like solutions, because they correspond to the saddle points of free energy $\mathcal{F}$ as functions of two skyrmion parameters, $R$ and $\delta$. Namely, at these points $\mathcal{F}[R,\delta]$ has a minimum as a function of $\delta$ and a maximum as a function of $R$.} represent skyrmion metastable states that should evolve to a certain stable state (some local minimum of $\Delta\mathcal{F}[R]$), but have longer lifetimes than any other intermediate state.

In Fig.~\ref{fig:FviaR}(b) we show the numerical solutions of Eq.~\eqref{eq:ELE_theta_coax} (solid curves) in comparison with the instances of  the ansatz (dashed curves) for the five extrema of $\Delta\mathcal{F}[R]$ in Fig.~\ref{fig:FviaR}(a). In order to plot $\theta_{R,\delta,\gamma}(r)$ we use $R$ and $\delta$ found by minimization of the free energy. There is a very good agreement between the numerical solution and the ansatz. We checked that such an agreement is a general situation provided $\gamma{\lesssim}1$.

The solutions with positive chirality shown in Fig.~\ref{fig:FviaR} have an interesting feature. The magnetization is parallel to $z$ axis not only at the origin and the infinity but also at two intermediate distances, see  Fig.~\ref{fig:FviaR}(b). This is related to the fact that at small distances from the origin the magnitude of the skyrmion angle is larger than $\pi$, whereas at large distances the angle becomes negative. Both features arise because the spatial dependence of the skyrmion angle at small and large distances is controlled by the vortex solution \eqref{eq:thetaV}, see dotted curves in Fig.~\ref{fig:FviaR}(b).

By setting different initial magnetizations in micromagnetic modeling, we have managed to observe all three stable skyrmion profiles for $\epsilon{=}0.3$ and $\gamma{=}0.522$. They are shown as color-plot insets in Fig.~\ref{fig:FviaR}. 
For three stable skyrmions we have also extracted a dependence of the skyrmion angle on the distance $r$. They are shown by
points in Fig.~\ref{fig:FviaR}(b). We emphasize that the 
numerical solution of Eq.~\eqref{eq:ELE_theta_coax}, ansatz~\eqref{eq:ansatz}, and the result of the micromagnetic simulations match remarkably well.

\begin{figure}[!t]
	\includegraphics[width=3.5in]{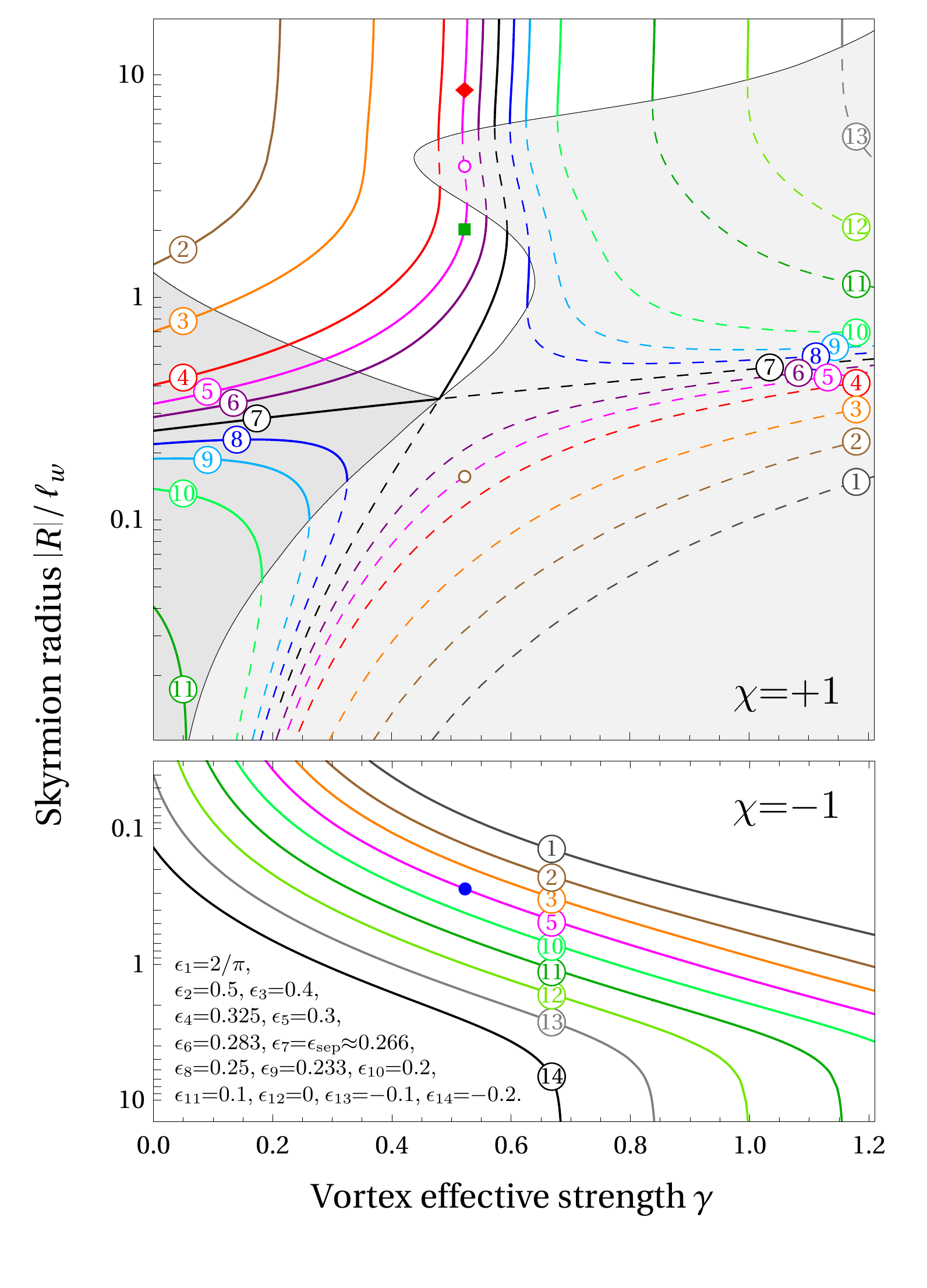}	
	\caption{Dependence of skyrmion radius~$|R|/\ell_w$ on the effective vortex strength~$\gamma$ for several values of DMI parameter $\epsilon$ in semi-log scale for chiralities~$\chi=\pm1$. The solid and dashed curves correspond to the minima and maxima of $\Delta\mathcal{F}[R]$, respectively. The area of unstable coaxial states and the area of metastable states are marked by the darker and lighter gray fillings, respectively. The values of $\epsilon$ used for the curves with the corresponding numbers from 1 to 14 are provided in the lower panel.
	The blue disk, green square, and red diamond correspond to the stable skyrmions in Fig.~\protect\ref{fig:FviaR}, while the brown and magenta circles show the solutions corresponding to maxima of $\Delta\mathcal{F}[R]$ there.}
	\label{fig:diagram}
\end{figure}

With the change of parameters $\epsilon$ and $\gamma$, which depend on a particular heterostructure, the number of extrema in $\Delta\mathcal{F}[R]$ may vary. To characterize these extrema at different $\epsilon$ and $\gamma$, we show in Fig.~\ref{fig:diagram} semi-log dependencies of $R$ on $\gamma$ for several values of $\epsilon$. The solid and dashed curves on the plane $(\gamma,R)$ 
correspond to the minima and maxima of $\Delta\mathcal{F}[R]$, respectively. The area of the saddle-like metastable states 
are marked by the lighter gray filling in Fig.~\ref{fig:diagram}.
 
The numerical solution of Eq.~\eqref{eq:ELE_theta_coax}
and ansatz~\eqref{eq:ansatz} give stable skyrmions provided their center is pinned to the center of the Pearl vortex. However, these states can be unstable with respect to the shifting of skyrmion center from the center of the vortex. Indeed, as shown in Ref.~\cite{Andriyakhina2021}, a skyrmion-vortex pair can be stable when the skyrmion is located at some finite distance~$a$ from the Pearl vortex. To determine the stability of the coaxial configuration we shift the skyrmion from the center to the infinitesimally small distance $a{\to}0$ and compare the free energy of the shifted and coaxial configurations. The corresponding analysis~\cite{new-JETP-letters-2022} showed that all coaxial skyrmion-vortex states with chirality $\chi{=}{+}1$ for $\epsilon{<}\epsilon_{\rm cr}{\approx}0.49$ and $\gamma{<}\gamma_{\rm cr}(\epsilon)$ are unstable and the skyrmion repulses from the vortex. In Fig.~\ref{fig:diagram} we mark the area of the unstable coaxial configurations by the darker gray filling. The diagram in Fig.~\ref{fig:diagram} has several interesting features.

 Firstly, all curves for skyrmions of chirality~$\chi{=}{+}1$ (the upper panel) are located in quadrants which are produced on the plane $(\gamma,R)$ by the curve at $\epsilon{=}\epsilon_{\rm sep}{\approx}0.266$ (black line \#7). For $\epsilon{<}\epsilon_{\rm sep}$ curves $R(\gamma)$ are located in the left-bottom and right-top quadrants, while the curves for $\epsilon{>}\epsilon_{\rm sep}$ are situated in the right-bottom and left-top quadrants. We emphasize that even for $\epsilon{\leqslant}0$ there are 
 skyrmions with~$\chi{=}{+}1$ in the right-top quadrant (see curves \#12 and \#13). 

Secondly, for 
$0.25{\lesssim}\epsilon{\lesssim}0.35$
there are values of $\gamma$ where two skyrmions of chirality $\chi{=}{+}1$ may exist. Recall that such situation was shown by the solid curve in Fig.~\ref{fig:FviaR}(a), where two minima of $\Delta\mathcal{F}[R]$ corresponding to positive chirality are shown.

Thirdly, for each pair of $\epsilon$ and $\gamma$ there is a skyrmion of chirality~${\chi{=}{-}1}$ with a certain radius~$|R|$, see the lower panel of Fig.~\ref{fig:diagram}. However, it should be emphasized, that for small $\gamma$ the radius of skyrmion for $\epsilon{>}0$ appears to be extremely small, $R{\lll}\ell_{w}$, that is beyond applicability of the free energy \eqref{eq:MagFe}. That is why we do not plot solutions with such small radii in Fig.~\ref{fig:diagram}. 
As distinctly seen from Fig.~\ref{fig:diagram}, the radius~$|R|$ for the skyrmions with~${\chi{=}{-}1}$ monotonically increases with the growth of~$\gamma$.

Finally, for each $\epsilon$ and for both chiralities $\chi{=}{\pm}1$ there is a critical value $\gamma_\pm(\epsilon)$. When $\gamma$ increases close to $\gamma_\pm$, the skyrmion radius significantly grows. For $|R|{\gg}\ell_{w}$ the free energy can be calculated in the 
leading approximation,
\begin{equation}
	\label{eq:gamma_pm}
\dfrac{\Delta\mathcal{F}[R]}{8\pi Ad_f}\approx
	(1\mp\epsilon\pi/2)|R|/\ell_{w} -\gamma \ell_{w}^{-2}\int_0^{|R|}d r\, r b_z(r).
\end{equation} 
The first term implies the energy of the domain wall, separating interior of the skyrmion from its exterior, while the second term comes from the energy of the inner area of skyrmion where magnetization, $m_z{\approx}{-}1$, is directed opposite to the magnetization in the ferromagnetic state, $m_z{\approx}{+}1$. For 
${\ell_{w}{\ll}|R|{\ll}\lambda}$, we can approximate $b_z(r){\approx}\ell_w/r$, and estimate the second term in Eq.~\eqref{eq:gamma_pm} as $\gamma|R|/\ell_{w}$. Therefore, the critical value of~$\gamma$ can be estimated as ${\gamma_\pm(\epsilon){\approx}1{\mp}\epsilon\pi/2}{>}0$. If $\gamma
{\gtrsim}\gamma_\pm$, the skyrmion radius becomes comparable or larger than the Pearl length, $|R|{\gtrsim}\lambda{\gg}\ell_{w}$ and the minimum of $\Delta\mathcal{F}[R]$ is determined by the relation
${ |R| b_z\big(|R|\big){\approx}\ell_{w}\gamma_\pm/\gamma}$.
For $\gamma{\gg}\gamma_\pm$, one can  
approximate the magnetic field of the vortex as  
${b_z\big(|R|\big)\approx4\ell_{w}\lambda^2/|R|^3}$ at $|R|{\gg}\lambda$. Therefore, the minimum of $\Delta\mathcal{F}[R]$ is achieved at ${|R|{\approx}2\lambda\sqrt{\gamma/\gamma_\pm}}$.

\noindent\textsf{\color{blue}Discussion.} The most interesting effects are predicted to occur in the range $\epsilon{=}0.25{\div}0.45$ and $\gamma{=}0.3{\div}0.7$, see Fig. \ref{fig:diagram}. However, in thus far experimentally available SF heterostructures \cite{Metaxas2007,Sampaio2013,Romming2013,Ryu2014,Romming2015,MoreauLuchaire2016,Petrovic2021}, $\epsilon$ varies from $0.25$ to $0.45$, whereas  $\gamma{\lesssim}0.1$ due to large magnitude of the Pearl penetration length. Therefore, in order to observe predicted effects one needs to enlarge $\gamma$ by increasing $d_S$ as well as by using cleaner superconductors to reduce $\lambda_L$. 

Our results imply that a superconducting vortex stabilizes a N\'eel-type skyrmion in the absence of DMI as shown by the curve \#12 in Fig. \ref{fig:diagram} (for similar effect in the absence of vortex see Ref. \cite{Kuznetsov2022}).

We note that experiments are usually performed under an external out-of-plane magnetic field. Such a field can be readily incorporated into our approach \cite{elsewhere}. 
Also, we mention that the case of antivortex ($\gamma{<}0$) cannot be easily related to the case $\gamma{>}0$ by some symmetry transformation and, therefore, requires separate investigation \cite{elsewhere}.
Finally, our theory can be extended to skyrmions and vortices in confined geometries \cite{Rohart2013,Vadimov2018,Gonzalez2022},
skyrmion-vortex lattices \cite{Neto2022}, and more exotic topological spin textures \cite{Gobel2021}, such as bimerons \cite{Gobel2019} and antiferromagnetic skyrmions \cite{Barker2016}.

\noindent\textsf{\color{blue}Summary.} Using three complementary approaches, namely, the direct numerical solution of the Euler-Lagrange equation~\eqref{eq:ELE_theta_coax}, the free energy minimization with two-parameter ansatz~\eqref{eq:ansatz}, and micromagnetic simulations, we developed the theory of the magnetization profile for the coaxial configuration of the N\'eel-type skyrmion and the Pearl vortex in thin superconductor-chiral ferromagnetic heterostructures. We found that the inhomogeneous magnetic field of the Pearl vortex significantly influences the skyrmion profile leading to drastic enhancement of the skyrmion's radius and inversion of the skyrmion's chirality. Both effects are controlled by
dimensionless magnetic field strength $\gamma$, proportional to the superconductor's thickness $d_S$. Such significant modification of magnetization profile of skyrmion in the presence of vortex can affect Majorana modes localized at skyrmion-vortex pairs \cite{Chen2015,Yang2016,Gungordu2018,Mascot2019,Rex2019,Garnier2019,Rex2020,Gungordu2022}. 

\begin{acknowledgements}
\noindent\textsf{\color{blue}Acknowledgements.} We thank A.~Fraerman, M.~Kuznetsov, and M.~Shustin for useful discussions. The work of S.S.A., E.A.S., and I.S.B. was funded by the Russian Science Foundation under the Grant No.~21-42-04410. O.A.T. acknowledges the support by the Australian Research Council (Grant No. DP200101027); the Cooperative Research Project Program at the Research Institute of Electrical Communication, Tohoku University (Japan); and an NCMAS grant.
\end{acknowledgements}

\bibliography{biblio-SkV}

\begin{thebibliography}{50}
\bibitem [{els()}]{elsewhere}%
  \BibitemOpen
  \href@noop {} {}\bibinfo {howpublished} {The correspondent studies will be
  published elsewhere.}\BibitemShut {Stop}%
\end{thebibliography}

\end{document}